\documentclass[pra,aps,superscriptaddress,twocolumn,floatfix]{revtex4-1}

\usepackage{dsfont}
\usepackage{mathtext}

\usepackage{mathtools}
\usepackage[usenames,dvipsnames]{xcolor}
\usepackage{amsmath}
\usepackage{amssymb,mathrsfs}
\usepackage{graphicx}
\usepackage{epstopdf}

\usepackage{mathtext}

\usepackage{dcolumn}
\usepackage{color}
\usepackage{bm}

\usepackage{times}

\newtheorem{lemma}{Lemma}

\newcommand{\br}{{\bm r}}

\newcommand{\tb}{\tilde{b}}
\newcommand{\tH}{\tilde{H}}

\newcommand{\tpsi}{\tilde{\psi}}

\newcommand{\RNum}[1]{\uppercase\expandafter{\romannumeral #1\relax}}

\newcommand{\rtext}[1]{\textcolor{red}{{#1}}}
\newcommand{\btext}[1]{\textcolor{blue}{#1}}

\begin{document}

\title{Topological dipole Floquet solitons}

\author{Sergey K. Ivanov}
\affiliation{Moscow Institute of Physics and Technology, Institutsky
lane 9, Dolgoprudny, Moscow region, 141700, Russia}
\affiliation{Institute of Spectroscopy, Russian
Academy of Sciences, Troitsk, Moscow, 108840, Russia}

\author{Yaroslav V. Kartashov}
\affiliation{Institute of Spectroscopy, Russian
Academy of Sciences, Troitsk, Moscow, 108840, Russia}
\affiliation{ICFO-Institut de Ciencies Fotoniques, The Barcelona Institute of Science and Technology, 08860 Castelldefels (Barcelona), Spain}

\author{Matthias Heinrich}
\affiliation{Institute for Physics, University of Rostock, Albert-Einstein-Str. 23, 18059 Rostock, Germany}

\author{Alexander Szameit}
\affiliation{Institute for Physics, University of Rostock, Albert-Einstein-Str. 23, 18059 Rostock, Germany}

\author{Lluis Torner}
\affiliation{ICFO-Institut de Ciencies Fotoniques, The Barcelona Institute of Science and Technology, 08860 Castelldefels (Barcelona), Spain}
\affiliation{Universitat Politecnica de Catalunya, 08034, Barcelona, Spain}

\author{Vladimir V. Konotop}
\affiliation{Departamento de F\'isica and Centro de F\'isica Te\'orica e Computacional, Faculdade de Ci\^encias, Universidade de Lisboa, Campo Grande, Ed. C8, Lisboa 1749-016, Portugal}

\begin{abstract}
We theoretically introduce a new type of topological dipole solitons propagating in a Floquet topological insulator based on a kagome array of
helical waveguides. Such solitons bifurcate from two edge states belonging to different topological gaps and have bright envelopes of different
symmetries: fundamental for one component, and dipole for the other. The formation of dipole solitons is enabled by unique spectral features
of the kagome array which allow the simultaneous coexistence of two topological edge states from different gaps at the same boundary.
Notably, these states have equal and nearly vanishing group velocities as well as the same sign of the effective dispersion coefficients. We
derive envelope equations describing components of dipole solitons and demonstrate in full continuous simulations that such states indeed can
survive over hundreds of helix periods without any noticeable radiation into the bulk.
\end{abstract}



\maketitle


{PhySH Subject Headings:} Solitons; Topological insulators, Floquet insulators.

~

Topological insulators represent a new phase of matter
characterized by qualitatively different behavior of excitations in the
bulk and at the edge of these topologically nontrivial materials. The
phenomenology of topological insulators, originally developed in
solid state physics~\cite{R1,R2}, was extended to diverse areas of physics,
where it stimulated numerous experimental realizations, e.g. in
mechanics~\cite{R3,R4}, acoustics~\cite{R5,R6}, in atomic~\cite{R7,R8}, optoelectronic~\cite{R9,R10,R11},
and various photonic~\cite{R12,R13,R14,R15,R16,R17,R18,R19} systems. The significant progress made
in linear topological photonics is described in reviews~\cite{R20,R21,R22}, \rtext{while
the investigation of topological effects in nonlinear systems is still in its
infancy. In such systems evolution of the topological edge states
may be considerably affected by nonlinearity and a whole set of
novel phenomena, ranging from topologically-protected lasing to the
formation of so-called topological edge solitons, becomes possible~\cite{R23,R24,R25}}. Nonlinearity has been shown to impact the modulational
stability of topological edge states~\cite{R26,R27,R28}, the direction of topological
currents~\cite{R29}, the appearance of topologically nontrivial phases~\cite{R30,R31,R32}, and to lead to bistability~\cite{R33}. Furthermore, nonlinearity can give
rise to topological closed currents in the bulk of the photonic insulator~\cite{R34,R35} and induce a topological current at its edges~\cite{R36}. Nonlinear
hybridization of topological and bulk states was observed in~\cite{R37}, and
the valley Hall effect for vortices in nonlinear system was predicted in~\cite{R38}.

Nonlinearity allows the formation of edge solitons --- unique states
that exhibit topological protection and simultaneously feature a rich
variety of shapes and interactions. Edge solitons were predicted in
photonic Floquet insulators in continuous~\cite{R26,R34,R39,R40} and discrete~\cite{R41,R42,R43,R44} models, and in polariton microcavities~\cite{R28,R45,R46}. Their
counterparts in nontopological photonic graphene were observed in~\cite{R47}. Floquet Bragg solitons were reported in~\cite{R48}. Topological (non-Floquet) systems also allow the formation of Dirac~\cite{R49}, Bragg~\cite{R50},
and valley Hall~\cite{R51} edge solitons. Even though such states may in
principle be encountered in many physical systems, potentially
including Bose-Einstein condensates in time-modulated potentials~\cite{R52,R53}, only fundamental edge solitons with bell-shaped amplitude profiles have been reported to date. The only exception is Floquet
dark-bright states introduced in~\cite{R40}, where opposite signs of the
dispersion in two components dictate the dark structure of one of
them --- nevertheless still representing a fundamental state.

Unlike regular solitons that are rigorously defined, the
corresponding concept in nonlinear Floquet insulators refers
generically to the observation of localized states in nonlinear
topological insulators. Here, by a Floquet soliton (FS) we denote a
beam localized in the $(x,y)$-plane near the interface between
topologically different materials, which bears the following two
properties: it belongs to a nonlinear family bifurcating from the
respective linear Floquet-Bloch edge state, and in the weakly-nonlinear limit its envelope represents a conventional soliton solution
of the averaged nonlinear equation with constant coefficients. Due to
the broken transversal (in the $(x,y)$-plane) and longitudinal (along
the $z$-direction) translational symmetries, such states are
intrinsically nonstationary, undergoing small-scale oscillations that,
\rtext{as our numerical simulations reported below reveal}, render them effectively
metastable, thus they decay during evolution albeit remarkably slowly.
Thus, the envelopes of FSs analyzed here for a kagome Floquet
insulator are described by soliton-bearing coupled nonlinear
Schrödinger (NLS) equations with constant coefficients and they
remain localized during distances that dramatically exceed longitudinal lattice helical
period~\cite{R26,R35}.

The \textit{dipole} FSs introduced here are comprised of contributions
from \textit{different} topological gaps with envelopes of different
symmetries. For the existence of such solitons, the linear edge
states they bifurcate from must have equal group velocities and at
the same time experience equal \textit{signs} of the dispersions
(understood here in terms of the Floquet-Bloch spectrum). Then the
system sustains FSs where both components are \textit{bright}. The dipole
envelope of the weaker component in such two-dimensional (2D)
states is held in shape only by the nonlinear coupling to the stronger
fundamental component, as in nontopological vector dipole solitons
in uniform media~\cite{R54,R55,R56,R57}. Dipole FSs are hybrid objects that are
confined to the edge due to their topological nature, while the
nonlinear self-phase modulation enables their localization along the
edge. This is in contrast to conventional scalar 2D dipole solitons characterized by identical localization mechanism in two transverse
dimensions~\cite{R58,R59,R60,R61}.

\begin{figure*}[t] \centering
\includegraphics[width=\textwidth]{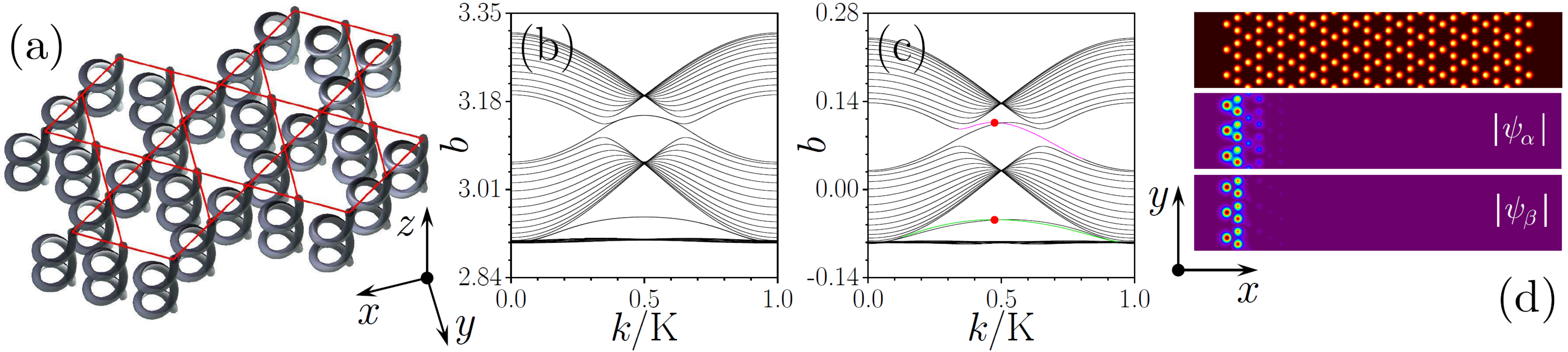}
\caption{(a)~Schematics of a kagome array of helical waveguides. (b)~Dependencies $b(k)$ showing three upper bands from the Bloch spectrum
for a finite array [see array profile in Fig.~\ref{Fig1}(d)] with straight waveguides $({{r}_{0}}=0)$. (c)~Quasi-propagation-constants $b(k)$ defined modulo
$\omega $ for a finite kagome array with helical channels at ${{r}_{0}}=0.6$, $T=8$. Red dots indicate edge states from different gaps with
equal velocities $\partial {{b}_{\alpha ,\beta }}/\partial k$. (d)~Three periods of finite kagome array (top) and linear Floquet eigenmodes ${{\psi
}_{\alpha ,\beta }}$ from the left edge (middle and bottom) at $z=0$, $k=0.472\operatorname{K}$ corresponding to the red dots in~(c). Here and
below $p=12$, $d=1.9$, $\sigma =0.4$.}
\label{Fig1}
\end{figure*}

The propagation of light along the $z$-axis of a helical kagome
array with focusing cubic nonlinearity is governed by the nonlinear
Schrödinger (NLS) equation for the dimensionless field amplitude
$\psi $:
\begin{equation}\label{NLSeq}
    \begin{split}
i\frac{\partial \psi }{\partial z}=-\frac{1}{2}{{\Delta }_{\bot }}\psi -
\mathcal{R}(\mathbf{r},z)\psi -{{\left| \psi \right|}^{2}}\psi \, .
    \end{split}
\end{equation}
Here $\mathbf{r}=\mathbf{i}x+\mathbf{j}y$ is the radius-vector in the
transverse plane, $x,y$ are the normalized transverse coordinates,
${{\Delta }_{\bot }}={{\partial }^{2}}/\partial {{x}^{2}}+{{\partial
}^{2}}/\partial {{y}^{2}}$; $z$ is the normalized propagation distance
and the refractive index profile is described by the function
$\mathcal{R}(\mathbf{r},z)=\mathcal{R}(\mathbf{r},z+T)=\mathcal{R}(
\mathbf{r}+L\mathbf{j},z)$. The array is comprised of identical
waveguides of width $\sigma $ placed in the nodes
${{\mathbf{r}}_{nm}}$ of the kagome grid
$\mathcal{R}(\mathbf{r},z)=p\sum\nolimits_{nm}{{{e}^{-{{[\mathbf{r}-
{{\mathbf{r}}_{nm}}-\mathbf{s}(z)]}^{2}}/{{\sigma }^{2}}}}}$, where $p$
is the array depth, and $\mathbf{s}(z)={{r}_{0}}(\sin (\omega z),\cos
(\omega z)-1)$ describes helical trajectory of each waveguide with
the Floquet period $T=2\pi /\omega $ and radius ${{r}_{0}}$ [Fig.~\ref{Fig1}(a)]. The $y$-period of such array is $L=2d$, where $d$ is the
separation between waveguides. To obtain edge states, we truncate
the array in the $x$-plane to form zigzag boundaries [Fig.~\ref{Fig1}(d)].
Typical parameters of such structures are $d\sim 1.9$ ($19\;\mu
\text{m}$ spacing), ${{r}_{0}}\sim 0.0-1.0$ (helix radius up
to $10 \;\mu\text{m}$), $p\sim 12$ (refractive index $\delta
n\sim 9\times {{10}^{-4}}$), $\sigma \sim 0.4$ ($4\;\mu
\text{m}$ wide waveguides), $T\sim 0-10$ (helix periods
up to $12\;\text{mm}$). We assume excitation at $\lambda
=800\;\text{nm}$. Arrays with such parameters can be readily created by
femtosecond laser inscription~\cite{R17}. \rtext{Notice that topological protection
of linear and nonlinear scalar modes in Floquet insulators with helical
channels and similar array parameters have been shown in Refs.~\cite{R26,R40}, therefore we expect the same degree of
protection also for the dipole vector states analyzed below.}

Linear eigenmodes of the helical array are Floquet-Bloch waves
$\psi (\mathbf{r},z)={{\phi }_{\nu k}}(\mathbf{r},z){{e}^{i{{b}_{\nu
k}}z}}$, where ${{\phi }_{\nu k}}(\mathbf{r},z)={{u}_{\nu k}}(\mathbf{r},z){{e}^{iky}}$ and the function ${{u}_{\nu
k}}(\mathbf{r},z)={{u}_{\nu k}}(\mathbf{r},z+T)={{u}_{\nu
k}}(\mathbf{r}+L\mathbf{j},z)$ is periodic along both $z$ and $y$ axes. Here $k$ denotes the Bloch momentum in the first Brillouin
zone $k\in [-\operatorname{K}/2,+K/2)$, where
$\operatorname{K}=2\pi /L$, $\nu $ is the mode index, while
${{b}_{\nu k}}\in [-\omega /2,+\omega /2)$ is a quasi-propagation-constant defined modulo $\omega $ and describing the phase
${{b}_{\nu k}}T$ accumulated by the Floquet-Bloch wave over one
$z$-period. A representative spectrum of a truncated kagome array
with straight waveguides (in this case, at ${{r}_{0}}=0$, $b$ is a
standard propagation constant) is presented in Fig.~\ref{Fig1}(b) (for brevity,
we omit subscripts in ${{b}_{\nu k}}$ in the figures). We show three
upper bands, the lowest of which is nearly flat. Two pairs of
degenerate edge states are clearly visible in the spectrum. For any
non-zero helix radius ${{r}_{0}}\ne 0$, the system becomes
topologically nontrivial as time-reversal symmetry is broken~\cite{R62,R63,R64}.
As a result, topological states occur on the left edge (we assign
indices $\nu =\alpha ,\beta $ to the ``top'' and ``bottom'' states) as
marked by magenta and green lines in Fig.~\ref{Fig1}(c). Representative
profiles of Floquet-Bloch waves from different gaps are shown in Fig.~\ref{Fig1}(d) (at $z=0$).

\begin{figure}[t] \centering
\includegraphics[width=0.48\textwidth]{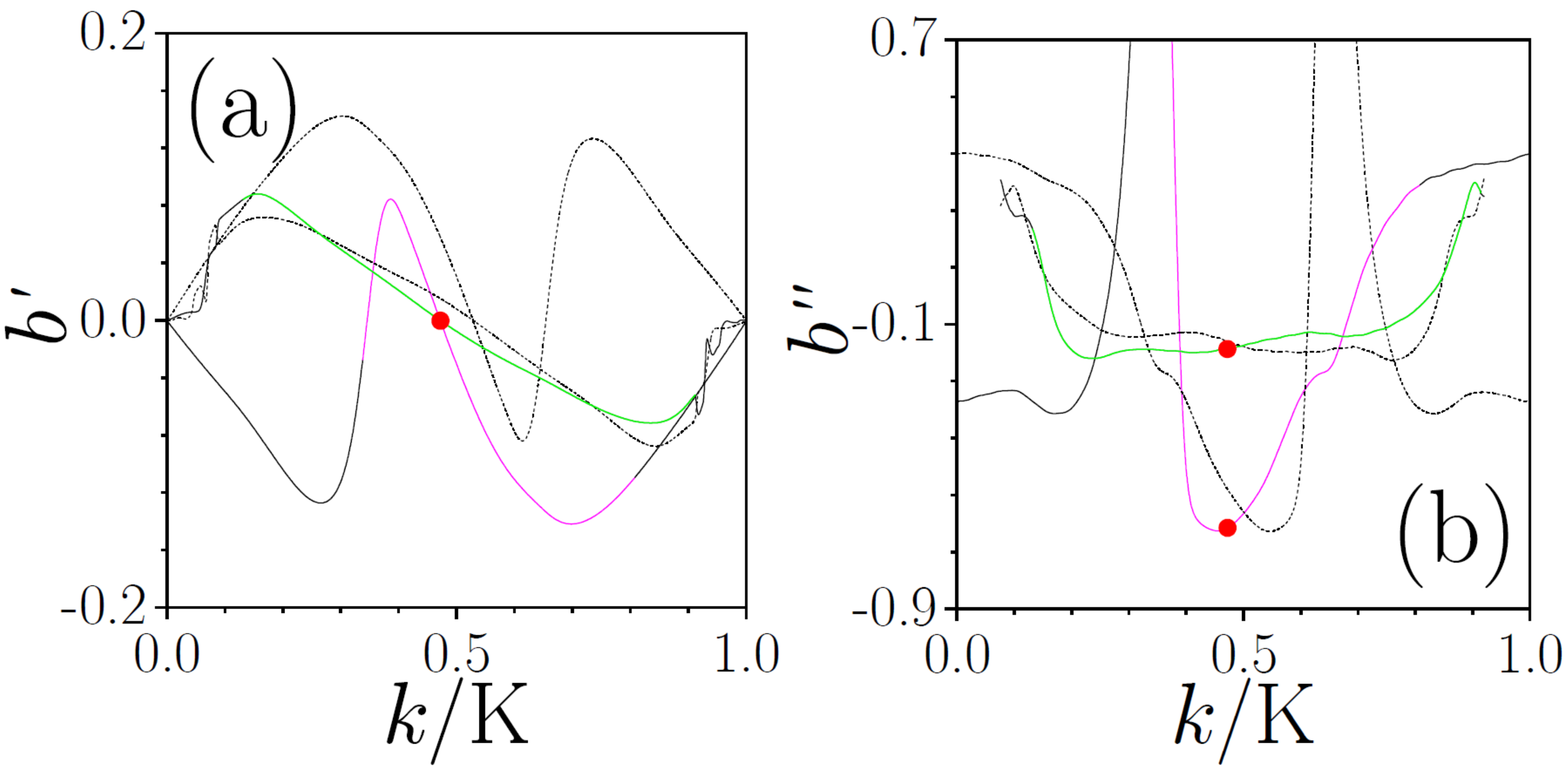}
\caption{Derivatives ${{b}'_{\nu }}$~(a) and ${{b}''_{\nu }}$~(b) for
the edge state branches. Solid (dashed) lines correspond to the
states from left (right) edges. Red dots indicate states from the left
edge with equal group velocities ${{b}''_{\alpha ,\beta }}=-0.00033$
and negative dispersion ${{b}''_{\alpha }}=-0.67331$, ${{b}''_{\beta
}}=-0.16827$ from which FSs bifurcate (see below). The color coding
for different branches is the same as in Fig.~\ref{Fig1}(c).}
\label{Fig2}
\end{figure}

A remarkable property of the kagome topological insulator is that
the derivatives ${{{b}'}_{\nu }}=\partial {{b}_{\nu k}}/\partial k$, defining
the group velocities ${{v}_{\nu }}=-{{{b}'}_{\nu }}$ of two topological
states coexisting at a given edge (see
\btext{Appendix A} for details) may coincide
for certain values of the Bloch momentum $k$ [see, e.g. the red dots
in Fig.~\ref{Fig2}(a)]. Importantly, the sign of the dispersion coefficients
${{{b}''}_{\nu }}={{\partial }^{2}}{{b}_{\nu k}}/\partial {{k}^{2}}$ for the
momentum corresponding to the red dots is likewise identical [Fig.
2(b)]. Coexistence of topological edge states with coinciding group
velocities ${{v}_{\alpha }}={{v}_{\beta }}$ and equal signs of the
effective diffraction in the underlying linear system is necessary for
the formation of multipole FSs as it allows for persistent nonlinearity-
mediated coupling. For our parameters, the group velocities coincide
at $k=0.472{K}$ (see Fig.~\ref{Fig2}).

To construct multipole FSs we focus on their bifurcation from the
linear Floquet-Bloch states ${{\psi }_{\alpha k}}$ and ${{\psi }_{\beta
k}}$. To this end we look for the solution in the form $\psi \approx
{{A}_{\alpha }}(Y,z){{\phi }_{\alpha k}}{{e}^{i{{b}_{^{\alpha
k}}}z}}+{{A}_{\beta }}(Y,z){{\phi }_{\beta k}}{{e}^{i{{b}_{\beta k}}z}}$,
where ${{A}_{\alpha ,\beta }}$ are the slowly-varying envelopes and
$Y=y-{{v}_{\alpha ,\beta }}z$ is the coordinate in the frame moving
with velocity ${{v}_{\alpha ,\beta }}=-{{{b}'}_{\alpha ,\beta }}$, identical
for both components. We adopt a multiscale expansion that shows that the envelopes ${{A}_{\alpha ,\beta }}$ are governed by the
coupled focusing NLS equations
(\btext{see Appendix A}):
\begin{equation}\label{NLSeq2}
    \begin{split}
    i\frac{\partial {{A}_{\alpha ,\beta }}}{\partial z}=&\frac{{{{{b}''}}_{\alpha ,\beta }}}{2}\frac{{{\partial }^{2}}{{A}_{\alpha ,\beta }}}{\partial {{Y}^{2}}} \\
    &\quad-({{\chi }_{\alpha ,\beta }}{{\left| {{A}_{\alpha ,\beta }} \right|}^{2}}+2{{\chi }_{\operatorname{x}}}{{\left| {{A}_{\beta ,\alpha }} \right|}^{2}}){{A}_{\alpha ,\beta }}\,,
    \end{split}
\end{equation}
where ${{\chi }_{\nu }}={{\langle ({{\left| {{\phi }_{\nu k}}
\right|}^{2}},{{\left| {{\phi }_{\nu k}} \right|}^{2}})\rangle }_{T}}$ and
${{\chi }_{\operatorname{x}}}={{\langle ({{\left| {{\phi }_{\alpha k}}
\right|}^{2}},{{\left| {{\phi }_{\beta k}} \right|}^{2}})\rangle }_{T}}$ are the
effective self- and cross-modulation coefficients, averaging over one
$z$-period is defined as ${{\langle g\rangle }_{T}}={{T}^{-
1}}\int_{0}^{T}{g(\mathbf{r},z)dz}$, and calculation of the inner
product
$(f,g)=\int_{S}{{{f}^{*}}(\mathbf{r},z)g(\mathbf{r},z)d\mathbf{r}}$ is
performed over the entire transverse array area $S$. Floquet-Bloch
states ${{\phi }_{\nu k}}$ are orthogonal and normalized at the same
instant $z$
(\btext{see Appendix A}): $({{\phi }_{\nu k}},{{\phi }_{{\nu }'k}})={{\delta }_{\nu
{\nu }'}}$. Note that the considerable difference between quasi-
propagation-constant mismatch $\delta {{b}_{k}}={{b}_{\alpha k}}-
{{b}_{\beta k}}\approx 0.15$ and frequency of periodic modulation
$(\omega \approx 0.78)$ ensures that coupling between the modes
is entirely non-resonant and therefore exclusively mediated by
nonlinearity. \rtext{Efficient nonlinear coupling between waves with
different momenta $k$ can only occur for a special ratio of propagation
constants of the involved topological states, which is not met in our system.}

\begin{figure}[t] \centering
\includegraphics[width=0.5\textwidth]{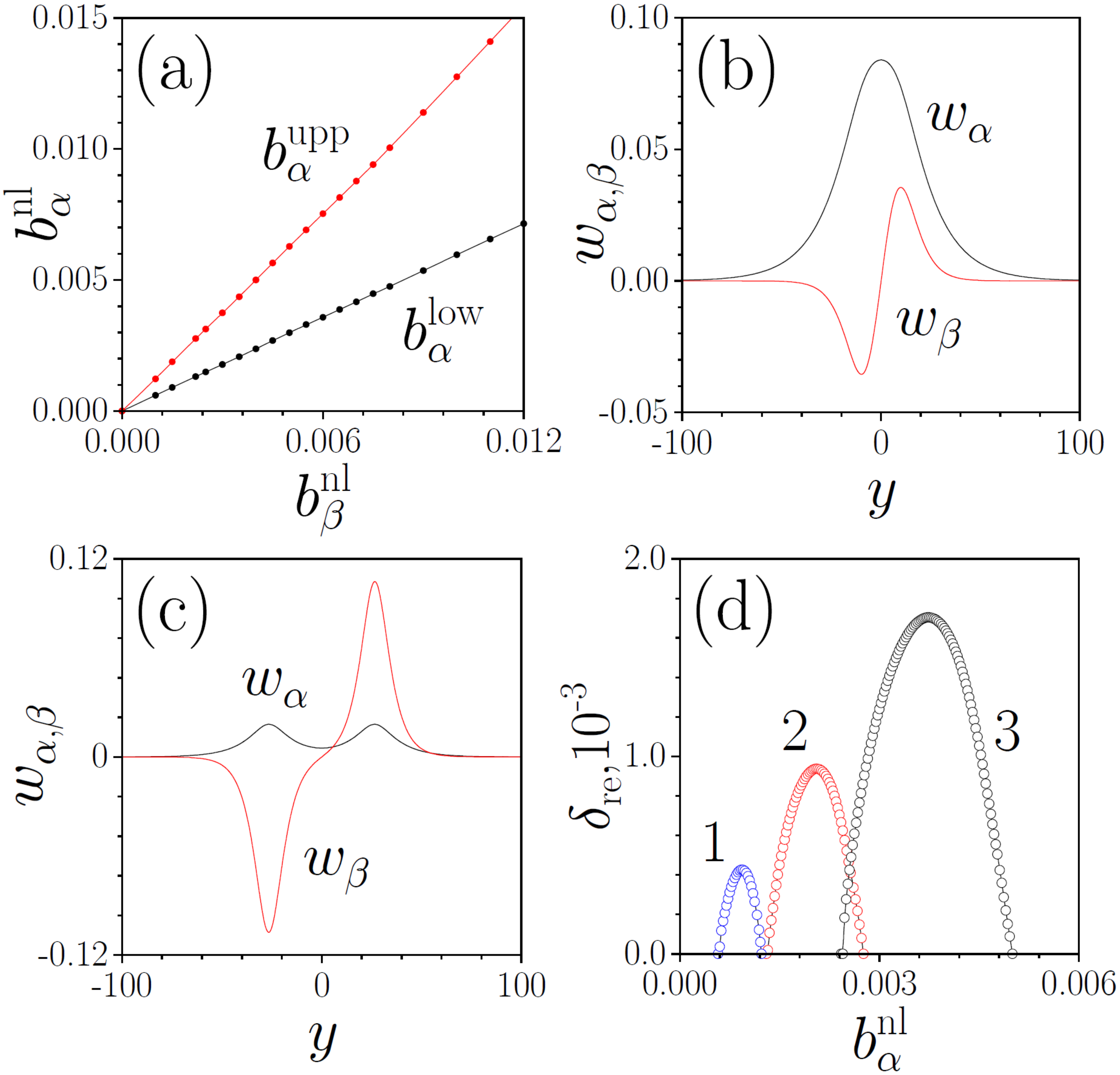}
\caption{(a)~Domain of dipole soliton existence on the $(b_{\alpha
}^{\operatorname{nl}},b_{\beta }^{\operatorname{nl}})$ plane. Dipole
soliton envelopes at $b_{\alpha }^{\operatorname{nl}}=0.0015$ (b)~and $b_{\alpha }^{\operatorname{nl}}=0.0027$ (c)~for $b_{\beta
}^{\operatorname{nl}}=0.0022$. (d) Maximal real part of perturbation
growth rate versus $b_{\alpha }^{\operatorname{nl}}$ at $b_{\beta
}^{\operatorname{nl}}=0.001$ (curve~1), $0.0022$ (curve~2), and
$0.004$ (curve~3). The parameters in the envelope equation at
$k=0.472{K}$ are ${{b}'_{\alpha }}={{b}'_{\beta }}=-
0.00033$, ${{b}''_{\alpha }}=-0.67331$, ${{b}''_{\beta }}=-0.16827$, ${{\chi }_{\alpha }}=0.31048$, ${{\chi }_{\beta }}=0.36011$,
${{\chi }_{\operatorname{x}}}=0.31973$.}
\label{Fig3}
\end{figure}

We are interested in bright dipole soliton solutions of Eqs.~(2) that
exist at ${{b}''_{\alpha }},{{b}''_{\beta }}<0$ [see Fig.~\ref{Fig2}(b)]. In such
states the bell-shaped $\alpha $ component prevents (by creating an
effective potential well via cross-phase modulation) out-of-phase
poles of the dipole $\beta $ component from splitting, leading to the
formation of stationary states. They can be found by the Newton
method in the form ${{A}_{\alpha ,\beta }}={{w}_{\alpha ,\beta
}}{{e}^{ib_{\alpha ,\beta }^{\operatorname{nl}}z}}$, where the
nonlinearity-induced phase shifts $b_{\alpha ,\beta
}^{\operatorname{nl}}$ should be sufficiently small (much smaller
than the quasi-propagation constants, the topological-gap width, and
longitudinal Brillouin zone $\omega $) to ensure that the profiles
${{w}_{\alpha ,\beta }}$ are broad and satisfy the assumption of slow
variation of the soliton profile. The properties of dipole solitons for
nonlinear and dispersion coefficients corresponding to the edge
states at $k=0.472\operatorname{K}$ are described in Fig.~\ref{Fig3}. For a
fixed $b_{\beta }^{\operatorname{nl}}$ dipole solitons exist for
$b_{\alpha }^{\operatorname{low}}\le b_{\alpha
}^{\operatorname{nl}}\le b_{\alpha }^{\operatorname{upp}}$. The
existence domain expands with $b_{\beta }^{\operatorname{nl}}$
[Fig.~\ref{Fig3}(a)]. Close to its lower border $b_{\alpha
}^{\operatorname{low}}$ the dipole $\beta $ component vanishes
and only fundamental $\alpha $ component remains, while at the
upper border $b_{\alpha }^{\operatorname{upp}}$ the soliton splits
into two states gradually separating as the amplitude of the $\alpha $
component vanishes. Representative profiles are shown in Fig.~\ref{Fig3}(b)
and~\ref{Fig3}(c). By substituting the perturbed envelope solitons ${{A}_{\nu
}}=({{w}_{\nu }}+{{\mu }_{\nu }}{{e}^{\delta z}}+\eta _{\nu
}^{*}{{e}^{{{\delta }^{*}}z}}){{e}^{ib_{\nu }^{\operatorname{nl}}z}}$,
where ${{\mu }_{\nu }},{{\eta }_{\nu }}\ll {{w}_{\nu }}$, into Eq.~(2) one
arrives at a linear eigenvalue problem
\btext{(see Appendix B)}, whose solution yields the
growth rate ${{\delta
}_{\operatorname{re}}}=\operatorname{Re}\delta $ for the most
unstable perturbation depicted in Fig.~\ref{Fig3}(d) as a function of $b_{\alpha
}^{\operatorname{nl}}$. The growth rate ${{\delta
}_{\operatorname{re}}}$ vanishes when $b_{\alpha
}^{\operatorname{nl}}\to b_{\alpha }^{\operatorname{low}},b_{\alpha
}^{\operatorname{upp}}$ and for the broad states considered here it
remains well below ${{10}^{-3}}$ implying that the characteristic
scale $1/{{\delta }_{\operatorname{re}}}$ of the instability
development exceeds hundreds of helix periods $T$.

\begin{figure*}[t] \centering
\includegraphics[width=0.8\textwidth]{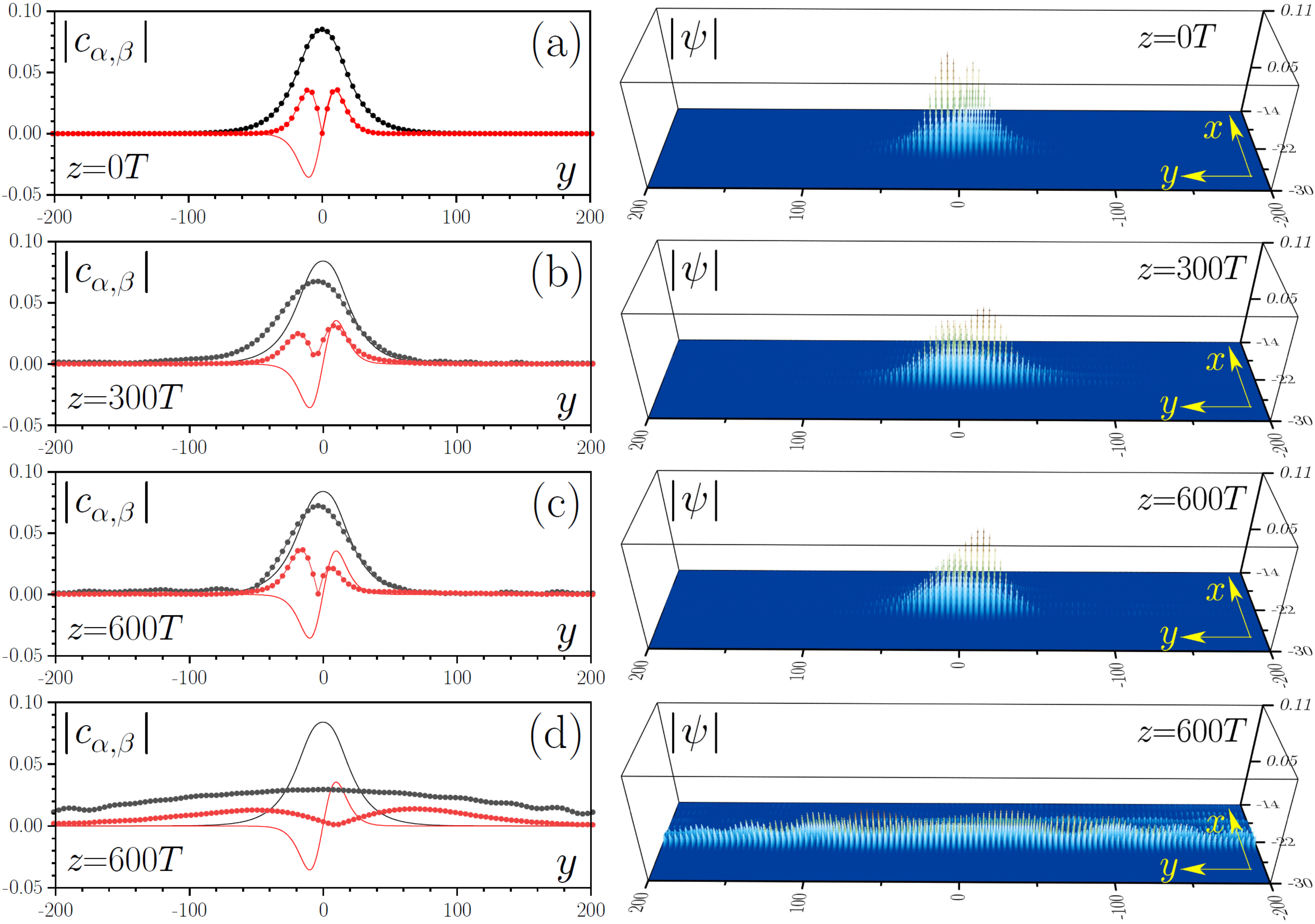}
\caption{Propagation of a dipole topological quasi-soliton in Eq.~(1) with envelope corresponding to $b_{\alpha }^{\operatorname{nl}}=0.0015$,
$b_{\beta }^{\operatorname{nl}}=0.0022$ in the helical kagome array in the nonlinear regime (a)-(c) and its diffraction in the linear regime (d).
Left column shows initial envelopes of two components (solid lines) and projections $\left| {{c}_{\alpha ,\beta }} \right|$ at different distances
(dots). The right column shows corresponding $\left| \psi \right|$ distributions.}
\label{Fig4}
\end{figure*}

To confirm the accuracy of the model~(2) and to confirm that
topological dipole solitons are observable experimentally, we
propagated Floquet-Bloch modes with exact dipole soliton
envelopes, obtained from Eq.~(2) for various $b_{\alpha ,\beta
}^{\operatorname{nl}}$ values, in the helical kagome array. Such
evolution is governed by the full 2D model~(1), which we solved with
a split-step FFT method. The input for Eq.~(1) was constructed as
$\psi ={{w}_{\alpha }}(Y){{\phi }_{\alpha k}}+{{w}_{\beta }}(Y){{\phi
}_{\beta k}}$. In the right column of Fig.~\ref{Fig4}(a)-(c) we show (with dots)
the modulus of the projections of the field $\psi $ on the linear
Floquet-Bloch modes: ${{c}_{\nu }}=\int_{mL-d}^{mL+d}{\phi _{\nu
k}^{*}(\mathbf{r},z)\psi (\mathbf{r},z)d\mathbf{r}}$ ($m\in \mathbb{Z}$
defines the $y$-period on which projection is calculated), and the
input envelopes ${{w}_{\alpha ,\beta }}$ (solid lines). The projections
${{c}_{\nu }}$ explicitly show that dipole soliton at all distances shown
contains contributions from two Floquet-Bloch states, whose
amplitudes remain practically unchanged and whose envelopes
remain mutually localized.

Propagation governed by~(1) confirms the metastability of the
dipole solitons, which survive over hundreds of helix periods even
when small-scale noise (up to 5\% in amplitude) is added into the
input field distributions. The rotation of the waveguides induces fast
$z$-oscillations of the soliton peak amplitude (a signature of its
Floquet nature) and causes very weak radiation, which nevertheless
does not destroy the dipole solitons at the considered distances. The
weak radiation becomes noticeable only at propagation distances
exceeding the ones shown here at least by one order of magnitude.
Metastability \rtext{[associated with very small, but nonzero growth rates
${{\delta }_{\operatorname{re}}}$ for perturbations of the envelope in 
Eq.~(2)]} results also in an extremely-slowly growth of the oscillations
of the two poles (peaks) of the dipole component (small input noise
only slightly affects phase of these oscillations), which nevertheless
do not cause splitting of the dipole state at least up to
$z<{{10}^{3}}T$. Splitting may occur, but at larger distances. The
right column of Fig.~\ref{Fig4}(a)-(c) illustrates the corresponding evolution of
the total field $\psi $. Since the group velocities of the two
components are close to zero, the soliton remains virtually locked in
place for the parameters chosen above, although for other helix
parameters we obtained slowly moving states. If nonlinearity is
switched off, wavepackets experience strong diffraction along the
array edge at similar propagation distances [Fig.~\ref{Fig4}(d)], an
observation that further confirms that the state from Fig.~\ref{Fig4}(a)-(c) is
sustained by nonlinearity.

\begin{figure*}[t] \centering
\includegraphics[width=0.8\textwidth]{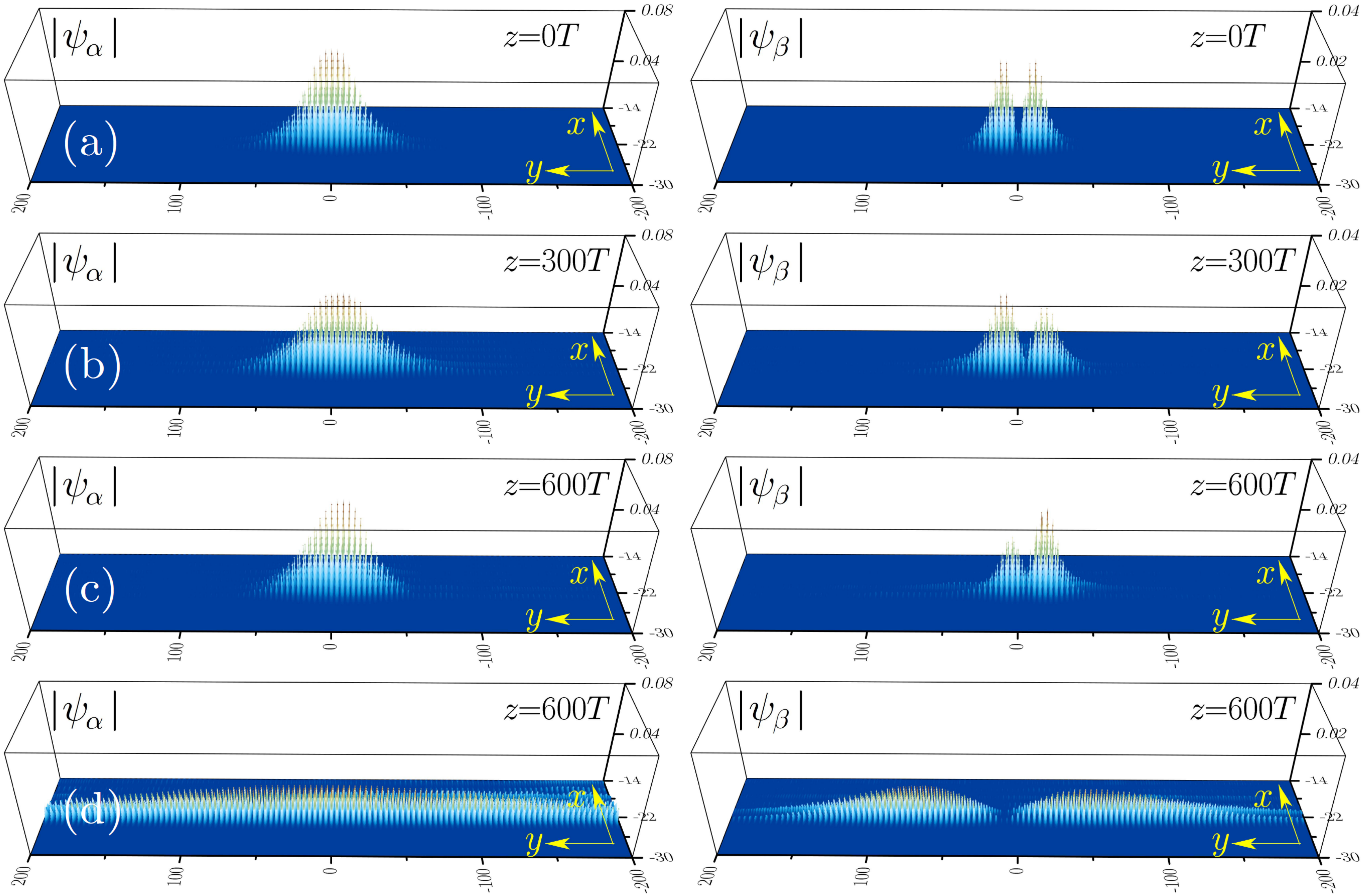}
\caption{Propagation of dipole topological quasi-soliton in equivalent vector Eq.~(3) in nonlinear medium (a)-(c) and its diffraction in linear regime
(d). Left column shows $\left| {{\psi }_{\alpha }} \right|$, while right column shows $\left| {{\psi }_{\beta }} \right|$. Parameters are the same as in
Fig.~\ref{Fig4}.}
\label{Fig5}
\end{figure*}

When the combination of two modes ${{\psi }_{\alpha }}$ and
${{\psi }_{\beta }}$ with different propagation constants \rtext{(i.e., a total field of the form $\psi \sim {{\psi
}_{\alpha }}+{{\psi }_{\beta }}$)} is substituted
into~(1), one can formally reduce it to two purely nonlinearly coupled
2D NLS equations \rtext{by collecting terms $\sim {{e}^{i{{b}_{\alpha
k}}z}},{{e}^{i{{b}_{\beta k}}z}}$ and dropping the oscillating terms $\sim
{{e}^{i({{b}_{\alpha k}}-{{b}_{\beta k}})z}}$} (thus, accounting only for self- and cross-phase modulation interactions \rtext{and skipping four-wave mixing terms}), without averaging over helix period $T$:
\begin{equation}\label{NLSeq3}
    \begin{split}
    i\frac{\partial {{\psi }_{\alpha ,\beta }}}{\partial z}=&-\frac{1}{2}{{\Delta }_{\bot }}{{\psi }_{\alpha ,\beta }}-\mathcal{R}(\mathbf{r},z){{\psi }_{\alpha ,\beta }} \\
    &\qquad-({{\left| {{\psi }_{\alpha ,\beta }} \right|}^{2}}+2{{\left| {{\psi }_{\beta ,\alpha }} \right|}^{2}}){{\psi }_{\alpha ,\beta }}\,.
    \end{split}
\end{equation}
The advantage of such a reduction is that~(3) allows to follow the
evolution of each component. This reduction is partially justified due
to rapid variation of phase difference $({{b}_{\alpha k}}-{{b}_{\beta
k}})z$ between modes, but it has to be tested numerically because
the scale ${{({{b}_{\alpha k}}-{{b}_{\beta k}})}^{-1}}>T$ is not the
smallest one in the Floquet system. The model~(3) can be also
directly derived for two waves with different
polarizations/wavelengths. The propagation of the dipole FS in the
vector model~(3) with a helical kagome array is illustrated in Fig.~\ref{Fig5}(a)-(c). Indeed, it shows metastable propagation of the dipole
soliton, qualitatively similar to the dynamics encountered in the
scalar model (Fig.~\ref{Fig4}). Also, the aforementioned oscillations of the
dipole component at the equivalent distances closely match the
oscillations of the corresponding projections in Fig.~\ref{Fig4} (notice the
different direction of the $y$-axis in panels with projections). As in
the scalar model, switching-off nonlinearity causes strong diffraction
(see
\btext{Appendix C} for the evolution of the peak amplitudes in the linear and
nonlinear cases). The remarkable similarity between the dynamics
in the scalar model~(1) and in the vector model~(3) shows that the
periodic modulation of the array does not introduce any linear
coupling of the involved modes.

In conclusion, we uncovered a new type of topological dipole
FS, which is constructed using envelopes featuring the different
symmetries imposed on two edge states from different topological
gaps exhibiting equal group velocities. The solitonic nature of the
wavepackets is consistent with their bifurcation from the linear
Floquet-Bloch eigenstates at small amplitudes and by the
preservation of their shape over extremely long propagation
distances. Our prediction has broad implications, as dipole solitons
can be observed for other types of Floquet insulators featuring at
least two topological gaps, such as, e.g., Floquet Lieb insulators. It is
plausible that more complex multicomponent solitons of non-
fundamental nature may be also found. Finally, we anticipate that
the reported results may be relevant for polaritonic and atomic
nonlinear systems, where topological edge solitons can be
sustained by different physical mechanisms.

\begin{acknowledgments}
Y.V.K. and S.K.I. acknowledge funding
of this study by RFBR and DFG according to the research project
no. 18-502-12080. A.S. acknowledges funding from the Deutsche
Forschungsgemeinschaft (grants BL 574/13-1, SZ 276/19-1, SZ
276/20-1). Y.V.K. and L.T. acknowledge support from the
Government of Spain (Severo Ochoa CEX2019-000910-S),
Fundaci\'o Cellex, Fundaci\'o Mir-Puig, Generalitat de Catalunya
(CERCA). V.V.K. acknowledges financial support from the
Portuguese Foundation for Science and Technology (FCT) under
Contract no. UIDB/00618/2020.
\end{acknowledgments}

\appendix

\section{Derivation of model~(2) from the main text}

Here we provide the detailed derivation of the coupled mode model, equation~(2) of the main text (see also~\cite{R40}).
We start with the model~(1) from the main text rewritten as follows
\begin{eqnarray}
\label{NLS}
i\frac{\partial\psi}{\partial z}= H_0(\br,z)\psi-|\psi|^2\psi, 
\end{eqnarray}
where $\br={\bf i} x+{\bf j}y$,
\begin{eqnarray} 
\quad H_0=-\frac{1}{2}\nabla^2
-\mathcal{R}(\br, z)
\end{eqnarray}
and the following properties hold
\begin{eqnarray}
\mathcal{R}(\br, z)=\mathcal{R}(\br, z+T)=\mathcal{R}(\br+L{\bf j}, z) 
\end{eqnarray}
with $T={2\pi}/{\omega}$, and other notations from the main text.

\subsection{The linear problem}

Consider the linear problem  
\begin{eqnarray}
\label{eq:linear}
i\frac{\partial\tpsi}{\partial z}= H_0\tpsi
\end{eqnarray}
(hereafter, tildes stand to distinguish  solutions of the linear problem from their nonlinear counterparts, i.e., $\tpsi$ is the linear limit of $\psi$).  A general Floquet-Bloch state (FBS) $\tpsi(\br,z)$   satisfies the Floquet (with respect to $z$) and Bloch (with respect to $y$) theorems and allows the representation 
\begin{equation}
\label{FBS}
\tpsi_{\nu k}(\br,z)=\phi_{\nu k}(\br,z)e^{i\tb_\nu(k)z}  =u_{\nu k}(\br,z)e^{iky+i\tb_\nu(k)z},  
\end{equation}
where $k\in[-K/2,K/2]$ with $K=2\pi/L$ is the Bloch vector along $y-$direction, $\tb_\nu(k)\in [-\omega/2, \omega/2]$, and
\begin{eqnarray}
u_{\nu k}(y,z)=u_{\nu k}(y+L,z)=u_{\nu k}(y,z+T)
\end{eqnarray}
(to abbreviate  notations we do not show $x-$dependence of $u_{\nu k}$ explicitly). The index $\nu$ stands either for a spatial band or for a topological branch connecting the neighbour gaps at a given edge.  

Now equation (\ref{eq:linear}) can be rewritten in terms of the functions $\phi_{\nu k}(y,z)$ and  $u_{\nu k}(y,z)$:
\begin{equation}
\label{eq:linear_phi}
i\frac{\partial \phi_{\nu k}}{\partial z}-\tb_\nu(k)\phi_{\nu k}= H_0\phi_{\nu k}
\end{equation}
and
\begin{equation}
\label{eq:linear_u}
i\frac{\partial u_{\nu k}}{\partial z}-\tb_\nu(k)u_{\nu k}= H_ku_{\nu k}\,,
\end{equation}
where
\begin{equation}
H_k=\frac{1}{2}\left(\frac{1}{i}\frac{\partial}{\partial y}+k\right)^2 -\frac{1}{2}\frac{\partial^2}{\partial x^2}+ \mathcal{R}(\br,z).
\end{equation} 
 
Let the lattice has dimensions defined by $x\in [-\ell_x,\ell_x] $ and $y\in [-\ell_y,\ell_y]$. Let also $\psi$ is subject to the cyclic boundary conditions with respect to $y$ and zero boundary conditions with respect to $x$ (these conditions were also used in numerical simulations):
\begin{align}
    \psi(\br,z)=\psi(\br+2\ell_y{\bf j},z), \quad \psi(\pm \ell_x{\bf i}+y{\bf j},z)=0. 
\end{align}
Respectively, $S=[-\ell_x,\ell_x]\times [-\ell_y,\ell_y]$ is the total area of the lattice. We are interested in the limit where $\ell_x,\ell_y\gg L$ (formally $\ell_x,\ell_y\to\infty$).
 
Define of the inner products  
\begin{equation}
\label{inner}
(f(\cdot,z),g(\cdot,z)):=\int_{S}f^*(\br,z) g(\br,z)d\br 
\end{equation}
and the $T-$average
 \begin{eqnarray}
 \label{ave_defin}
 \langle f\rangle_T:=\frac{1}{T}\int_{0}^{T}f(\br,z)dz.
 \end{eqnarray}
 We emphasize that the inner product in (\ref{inner}) is considered between two functions at the same instant. Below we use only such products and therefore drop the explicit argument $z$, writing $(f(\cdot,z),g(\cdot,z))=(f,g)$.

The following simple properties hold:

\begin{lemma}
 {The spectrum $\tb_\nu(k)$ is real.}  
\end{lemma} 
 
 \noindent
 {\em Proof:} Using (\ref{eq:linear_phi}) compute
 	\begin{eqnarray*}
 	 	i\left(\phi_{\nu k},\frac{\partial \phi_{\nu k}}{\partial z}\right)-\tb_\nu(k)(\phi_{\nu k},\phi_{\nu k})= (\phi_{\nu k},H_0\phi_{\nu k}).
 	 \end{eqnarray*}
  The complex conjugate of this equation reads  
 	 	\begin{eqnarray*}
 	 	-i\left(\frac{\partial \phi_{\nu k}}{\partial z},\phi_{\nu k}\right)-\tb_\nu^*(k)(\phi_{\nu k},\phi_{\nu k})= (\phi_{\nu k},H_0\phi_{\nu k}),
 	\end{eqnarray*}
 	where we used that $H_0$ is Hermitian in the Hilbert space with the inner product (\ref{inner}). 
 Subtracting one of this equation from another and applying the $T$-averaging we obtain
 \begin{equation}  
 i\Big\langle \frac{\partial}{\partial z} (\phi_{\nu k},\phi_{\nu k})\Big\rangle_T=  [\tb_\nu(k)-\tb_\nu^*(k)]\langle (\phi_{\nu k},\phi_{\nu k})\rangle_T\,.
 \end{equation}
 By the $T$-periodicity of $\phi_{\nu k}$, the l.h.s. is zero, and hence $\tb_\nu(k)=\tb_\nu^*(k)$. \quad $\Box$
 
 \begin{lemma}
 \label{lemma-ort}
 Non-degenerate states $\phi_{\nu' k'}(\br,z)$ and $\phi_{\nu k}(\br,z)$, with $(\nu,k)\neq (\nu',k')$, considered at the same instant $z$, are orthogonal  
 for all $z\geq0$.
  \end{lemma}
 
 \noindent
 {\em Proof:} Using (\ref{eq:linear_phi}) compute  
 \begin{equation}
 	\label{aux_phi_1}
 	i\left(\phi_{\nu' k'},\frac{\partial \phi_{\nu k}}{\partial z}\right)-\tb_\nu(k)(\phi_{\nu' k'},\phi_{\nu k})= (\phi_{\nu' k'},H_0\phi_{\nu k})
 \end{equation}
 and 
 \begin{equation}
 	i\left(\phi_{\nu k},\frac{\partial \phi_{\nu' k'}}{\partial z}\right)-\tb_{\nu'}(k')(\phi_{\nu k},\phi_{\nu' k'})= (\phi_{\nu k},H_0\phi_{\nu' k'}).
 \end{equation}
The complex conjugation of the last equation reads
\begin{equation}
	\label{aux_phi_3}
	-i\left(\frac{\partial \phi_{\nu' k'}}{\partial z},\phi_{\nu k}\right)-\tb_{\nu'}(k')(\phi_{\nu' k'},\phi_{\nu k})= (\phi_{\nu' k'}, H_0\phi_{\nu k}).
\end{equation}
Subtracting (\ref{aux_phi_3}) from (\ref{aux_phi_1}), we obtain
\begin{equation}
\label{aux_phi_4}
i\frac{d}{dz}\left(\phi_{\nu' k'},\phi_{\nu k}\right)+[\tb_{\nu}(k)-\tb_{\nu'}(k')](\phi_{\nu' k'},\phi_{\nu k})=0.
\end{equation}
This is a first order ODE for the inner product 
$\left(\phi_{\nu' k'},\phi_{\nu k}\right)$ as a function of $z$, and by assumption of non-degeneracy $\tb_{\nu}(k)\neq \tb_{\nu'}(k')$. Thus, if 
\begin{eqnarray}
\label{orthogonal}
 \left(\phi_{\nu' k'},\phi_{\nu k}\right)=0
\end{eqnarray}
at $z=0$,  then this property holds for  all $z\geq 0$. 

To complete the proof we notice that for a given fixed $z$, in particular for $z=0$ the Hamiltonian $H_0$ is Hermitian, and thus (\ref{orthogonal}) holds. \quad $\Box$  

For the next consideration we notice the property, valid for the states from the different bands (of branches) $\nu$ and $\nu'$ but equal Bloch wavenumbers $k=k'$:
\begin{align}
    \left(\phi_{\nu' k}(\cdot,z),\phi_{\nu k}(\cdot,z)\right)=\left(u_{\nu' k}(\cdot,z),u_{\nu k}(\cdot,z)\right).
\end{align}
We emphasize that the states $u_{\nu' k}^*(\br,z)$ and $u_{\nu k}(\br,z)$ are considered here at the same instant $z$

By Lemma~\ref{lemma-ort}, for $\nu\neq\nu'$ and the same instant $z$, $\left(u_{\nu' k},u_{\nu k}\right)=0$, while for $\nu=\nu'$ this inner product is a constant (independent on $z$). Thus, we can impose the normalization
\begin{align}
\label{normal}
    \left(u_{\nu' k}(\cdot,z),u_{\nu k}(\cdot,z)\right)=\delta_{\nu\nu'}\,.
\end{align}
This condition will be used in what follows.

 
\subsection{The ${\bm k}\cdot{\bm p}$ perturbation theory}

Now we extend the standard ${\bm k}\cdot{\bm p}$ perturbation theory~\cite{R67} to the case of $z-$dependent topological FBSs. 
Let $\tpsi_{\nu k}(\br,z)$ be a topological FBS. Consider also a FBS belonging to the same branch $\nu$ but having a Bloch wavevector $k_1=k+\delta k$ where $\delta k$ is infinitesimal. Taylor expansion of the dispersion relation yields
\begin{equation}
\label{eq:om_expan}
\tb_{\nu}(k+\delta k)=\tb_{\nu}(k)+\tb_{\nu}^{\prime}(k)\delta k+\frac{1}{2}\tb_{\nu}^{\prime\prime}(k)(\delta k)^2+\cdots
\end{equation}
(here a prime stands for the derivative with respect to $k$). 
On the other hand, equation for $u_{\nu k_1 }(\br,z)$   can be rewritten as
\begin{equation}
\label{eq:linear_u_exp}
i\frac{\partial u_{\nu k_1}}{\partial z}-\tb_\nu(k_1)u_{\nu k_1}= H_ku_{\nu k'}+H^{(1)}  u_{\nu k_1}\delta k+\frac{1}{2} u_{\nu k_1} (\delta k)^2,
\end{equation}
where
\begin{equation}
\label{eq:H1}
H^{(1)}=\frac{1}{i}\frac{\partial}{\partial y}+k.
\end{equation}

Now we compute $\tb_\nu(k_1)$ and $u_{\nu k_1}$ perturbatively from (\ref{eq:om_expan}), (\ref{eq:linear_u_exp}) and (\ref{eq:H1}). To this end, we look for a solution in the form of the expansion
\begin{align}
\label{exp_u_main}
u_{\nu k_1}=u_{\nu k}+\delta k\, u_{\nu k}^{(1)}+(\delta k)^2 u_{\nu k}^{(2)}+\cdots \;,
\end{align}
where $u_{\nu k}^{(1,2)}$ can be expanded over the complete set of the states $u_{\nu k}$
\begin{eqnarray}
\label{eq:exp_u}
 u_{\nu k}^{(j)}=\sum_{\lambda} c_{\nu \lambda}^{(j)}(z) u_{\lambda k}\,, \qquad j=1,2
\end{eqnarray}
(notice that the expansion coefficients $c_{\nu \lambda}^{(j)}(z)$ are functions of $z$).

Several important comments are in order. 


First, like in the stationary case (see, e.g.,~\cite{R68}) it is enough to consider only projections of $u_{\nu k}^{(1,2)}$ on the eigenstates with the same Bloch vector $k$, and therefore the sum in (\ref{eq:exp_u}) is over band number only (this is confirmed by the self-consistency of the expansion). Also this is the reason why the expansion coefficients $c(z)$ are not labeled by the index $k$: they all correspond to the chosen $k$. 

Second.  Unlike in the stationary perturbation theory, where the sum excludes also the state $u_{\nu k}(\br,z)$ to which the  perturbation $u_{\nu k}^{(1,2)}(\br,z)$ is orthogonal, now we have to keep this term because the orthogonality (\ref{orthogonal}) is verified only for the same instants $z$, while two states considered at different instants, say at $z_1$ and $z_2$ such that $z_1\neq z_2$, are, generally speaking, non-orthogonal. Physically, this means that the mode $u_{\nu k}(\br,z)$ can be excited at $z=z_2$ even if at the instant $z=z_1$ it is zero. 

Third, the functions $u_{\nu k}^{(1,2)}(\br,z)$ are $T-$periodic along $z$. This implies that the expansion coefficients $c_{\nu \lambda}^{(j)}(z)$ are also periodic functions of $z$, i.e., 
\begin{eqnarray}
\label{periodicity}
c_{\nu \lambda}^{(j)}(z+T)=c_{\nu \lambda}^{(j)}(z).
\end{eqnarray} 

Substituting (\ref{eq:om_expan}), (\ref{exp_u_main}), and (\ref{eq:exp_u}) into (\ref{eq:linear_u_exp}), collecting terms up to $(\delta k)^2$ order, taking into account that the states $u_{\nu k}$ solve Eq.~(\ref{eq:linear_u}), and separating the orders  $\delta k$ and $(\delta k)^2$ we obtain
\begin{widetext}
\begin{eqnarray}
\label{order1}
i\sum_{\lambda}\dot{c}_{\nu \lambda}^{(1)}u_{\lambda k}-\sum_{\lambda}(\tb_{\nu}-\tb_{\lambda})c_{\nu \lambda}^{(1)}u_{\lambda k}-\tb_\nu^{\prime} u_{\nu k}=H^{(1)}u_{\nu k}\,,
\\
\label{order2}
i\sum_\lambda\dot{c}_{\nu \lambda}^{(2)}u_{\lambda k}-\sum_{\lambda}(\tb_{\nu}-\tb_{\lambda})c_{\nu \lambda}^{(2)}u_{\lambda k}-\tb_\nu^{\prime}\sum_{\lambda}c_{\nu \lambda}^{(1)}u_{\lambda k}- \frac{1}{2}\tb_\nu^{\prime\prime}u_{\nu k}=H^{(1)}\sum_{\lambda} c_{\nu \lambda}^{(1)}u_{\lambda k}+\frac{1}{2}u_{\nu k}\,,
\end{eqnarray}
\end{widetext}
where the overdot stands for the derivative with respect to $z$: $\dot{c}=dc/dz$.

Applying  $(u_{\nu k},\cdot)$ to (\ref{order1}) we obtain
\begin{eqnarray}
\label{eq:order1_1}
i\frac{dc_{\nu \nu}^{(1)}}{dz}-\tb_\nu^{\prime}=(u_{\nu k},H^{(1)}u_{\nu k}).
\end{eqnarray}
By the requirement (\ref{periodicity})  $\langle dc_{\nu\nu}^{(j)}/dz\rangle_T=0$ and hence
\begin{eqnarray}
\label{omega1_}
\tb_{\nu}^{\prime}(k)= -\langle  (u_{\nu k},H^{(1)}u_{\nu k})\rangle_T \,.
\end{eqnarray}
We rewrite this expression as
 \begin{equation}
\label{omega1}
v_{\nu}(k)=-\tb_{\nu}^{\prime}(k)=  \left\langle \left(\phi_{\nu k},\frac{1}{i}\frac{\partial}{\partial y}\phi_{\nu k}\right)\right\rangle_T 
\end{equation}
(bellow we argue that $v_{\nu}(k)$ represents the group velocity).
 
In the absence of resonances, the $T-$periodic solution of (\ref{eq:order1_1}) satisfying zero initial condition, $c_{\nu\nu}^{(1)}(0)=0$ reads
\begin{eqnarray}
\label{b1}
c_{\nu\nu}^{(1)}(z)=\frac 1i \int_{0}^{z}\left[h_{\nu\nu}(z')-\langle h_{\nu\nu}\rangle_T \right]   dz',
\end{eqnarray}
where we defined a $T-$periodic  function
\begin{align}
\label{period-h}
h_{\nu\lambda}(z+T)=h_{\nu\lambda }(z):= \left(\phi_{\nu k},\frac{1}{i}\frac{\partial}{\partial y}\phi_{\lambda k}\right),  
\end{align}
thus ensuring the equality
\begin{eqnarray}
 \int_{0}^{T}\left[h_{\nu\lambda}(z')-\langle h_{\nu\lambda}\rangle_T \right]dz'=0.
\end{eqnarray}

Applying  $(u_{\lambda k},\cdot)$ to (\ref{order1}) we obtain the differential equations for $c_{\nu\lambda}^{(1)}(z)$ ($\lambda \neq \nu$)  
\begin{eqnarray}
\label{ode_b1}
i\dot{c}_{\nu\lambda}^{(1)}-\Delta_{\nu\lambda} c_{\nu\lambda}^{(1)}=h_{\nu\lambda}^*(z),\quad \Delta_{\nu\lambda}=\tb_{\nu}-\tb_{\lambda}\,.
\end{eqnarray}
Recalling (\ref{period-h}) and using the Fourier expansion
\begin{eqnarray}
h_{\nu\lambda}^*(z)=\sum_{m} h_{\nu\lambda }^{(m)}e^{-i f_m z}, \qquad f_m=2\pi \frac{m}{T}.
\end{eqnarray}
the $T-$periodic solution of (\ref{ode_b1}) can be written as
 \begin{eqnarray}
 \label{b1_fin}
 c_{\nu\lambda}^{(1)} (z)=\sum_m\frac{h_{\nu\lambda}^{(m)}} {f_m-\Delta_{\nu\lambda}}  e^{-if_m z}.
 \end{eqnarray}
 We notice that generally speaking   $c_{\nu\lambda}^{(1)} (0)\neq 0$. 
  
 Now we turn to the equation appearing in the second order: 
\begin{equation}
\frac{1}{2}\tb_{\nu}^{\prime\prime}=
  i\dot{c}_{\nu\nu}^{(2)} -\tb_\nu^{\prime} c_{\nu\nu}^{(1)} -\sum_{\lambda} (u_{\nu k},H^{(1)}u_{\lambda k})c_{\nu\lambda}^{(1)}
  -\frac{1}{2}\,.
\end{equation}
Notice that $c_{\nu\lambda}^{(j)}$ does not depend on $y$ while $H^{(1)}$ acts on the functions of $y$.
Since $\tb_\nu^{\prime\prime}$ is a constant, and all coefficients $c_{\nu\nu'}^{(j)}$ are $T-$periodic, the easiest wave to obtain $\tb_\nu^{\prime\prime}$ is to perform $T-$averaging. This gives
\begin{eqnarray}
 \frac{1}{2}\tb_\nu^{\prime\prime}=-\frac{1}{2} -\sum_{\lambda\neq \nu}\left\langle (u_{\nu k},H^{(1)}u_{\lambda k})c_{\nu\lambda}^{(1)}\right\rangle_T\,,
\end{eqnarray}
where we have taken into account (\ref{omega1_}). 
Comparing this with (\ref{eq:om_expan}) and returning to the FBSs we obtain the final form of the dispersion of the group velocity (the effective diffraction coefficient)
\begin{equation}
\label{omega2}
 \tb_{\nu}^{\prime\prime}(k)=-1- 2\sum_{\lambda\neq \nu}\left\langle h_{\nu\lambda} c_{\nu\lambda}^{(1)}\right\rangle_T\,.
\end{equation}

 \subsection{Multiple-scale expansion}
 
 Now we apply the multiple-scale expansion to the nonlinear problem (\ref{NLS}). To this end, we use two sets of scaled variables
 \begin{eqnarray}
 (y_0,y_1,y_2...):=(y, \mu y, \mu^2 y,...),
 \\
  (z_0,z_1,z_2...):=(z, \mu z, \mu^2 z,...),
 \end{eqnarray}
 where $\mu\ll 1$ is a formal small parameter. The scaled variables are treated as independent. Respectively 
 we have
\begin{eqnarray}
  H_0&=&\tH_0+\mu \tH^{(1)}+\mu^2 \tH^{(2)},
 \\
 \tH^{(1)}&=& -\frac{\partial^2}{\partial y_0\partial y_1},
 \\
 \tH^{(2)}&=& -\frac{\partial^2}{\partial y_0\partial y_2}-\frac{1}{2}\frac{\partial^2}{\partial y_1^2},
\end{eqnarray} 
where $\tH_0$ is $H_0$ with the substitution $y\to y_0$. We also have 
\begin{eqnarray}
\frac{\partial }{\partial z}=\frac{\partial}{\partial z_0}+\mu \frac{\partial}{\partial z_1}+\mu^2\frac{\partial}{\partial z_2}+\cdots\;.
\end{eqnarray}

In this work we are interested in evolution of two nonlinearly coupled modes having the same Bloch vector $k$  but belonging to different branches denoted as $\nu=\alpha$ and $\nu=\beta$  (see Fig. 1 in the main text). Respectively, we look for a solution of (\ref{NLS}) in the form 
\begin{align}
\label{psi_expan}
\psi&= \mu \left[A_\alpha (y_1,z_1)\phi_{\alpha k} e^{i\tb_\alpha z}+A_\beta (y_1,z_1)\phi_{\beta k} e^{i\tb_\beta z} \right]
\nonumber \\ &+ \mu^2 \left[\phi_\alpha^{(1)}e^{i\tb_{\alpha} z}+\phi_\beta^{(1)}e^{i\tb_{\beta} z}\right] 
\nonumber \\ &
+\mu^3 \left[\phi_\alpha^{(2)}e^{i\tb_{\alpha} z}+\phi_\beta^{(2)}e^{i\tb_{\beta} z}\right] 
\nonumber \\ &+\cdots\;.
\end{align}
Here  $A_\alpha$ and $A_\beta$ are slowly varying envelopes of the states $\phi_{\alpha k}$ and $\phi_{\beta k}$; in the arguments of  $A_{\alpha,\beta}$ only the most rapid variables are indicated, e.g., $A(y_1,z_1)$ stands for $A(y_1,,y_2,...;z_1, z_2,...)$. To shorten notations further we do not show $k$ in the arguments of $\tb_\alpha$ and $\tb_\beta$. 

At each instant $z_0$, the second and third order corrections in (\ref{psi_expan}) can be expanded as follows
\begin{subequations}
\begin{eqnarray}
\phi_\alpha^{(j)} =\sum_\nu B_{\alpha\nu}^{(j)}(x_1,z_0)\phi_{\nu k}(z_0),
\\
\phi_\beta^{(j)} =\sum_\nu B_{\beta\nu}^{(j)}(x_1,z_0)\phi_{\nu k}(z_0).
\end{eqnarray}
\end{subequations}
Now we substitute (\ref{psi_expan}) into (\ref{NLS}) considered in the scaled variables. In the first order of $\mu$, the obtained equation is identically satisfied. 

 \subsubsection{Order $\mu^2$}

In the second order we obtain
\begin{widetext}
 \begin{eqnarray}
 \label{mu2order}
 	& &\left[i\frac{\partial A_{\alpha}}{\partial z_1}\phi_{\alpha k}+i\sum_\nu \frac{\partial B_{\alpha\nu}^{(1)} }{\partial z_0}\phi_{\nu k} +i\sum_\nu B_{\alpha\nu}^{(1)} \frac{\partial \phi_{\nu k}}{\partial z_0} -\tb_{\alpha}\sum_\nu B_{\alpha\nu}^{(1)} \phi_{\nu k}\right]e^{i\tb_{\alpha} z}
 	\nonumber \\
 	& &+\left[i\frac{\partial A_{\beta}}{\partial z_1}\phi_{\beta k}+i\sum_\nu \frac{\partial B_{\beta\nu}^{(1)} }{\partial z_0}\phi_{\nu k} +i\sum_\nu B_{\beta\nu}^{(1)} \frac{\partial \phi_{\nu k}}{\partial z_0} -\tb_{\beta}\sum_\nu B_{\beta\nu}^{(1)} \phi_{\nu k}\right]e^{i\tb_{\beta} z}
 	\nonumber \\
 	& &=\left[ \tH_0\sum_\nu B_{\alpha\nu}^{(1)} \phi_{\nu k}+\tH^{(1)}A_\alpha\phi_{\alpha k}\right]e^{i\tb_{\alpha} z}+\left[ \tH_0\sum_\nu B_{\beta\nu}^{(1)} \phi_{\nu k}+\tH^{(1)}A_\beta\phi_{\beta k}\right]e^{i\tb_{\beta} z}.
 \end{eqnarray}
Collecting the terms $\propto e^{i\tb_{\alpha} z}$ and $\propto e^{i\tb_{\beta} z}$ separately, using (\ref{eq:linear_phi}) with $H_0$ replaced by $\tH_0$ with $\nu=\alpha$ and $\nu=\beta$, and using the explicit expression for $\tH_1$ we rewrite  (\ref{mu2order}) in the form of two equations as follows 
 \begin{subequations}
 \begin{eqnarray}
  \label{mu2order_11}
  i\frac{\partial A_{\alpha}}{\partial z_1}\phi_{\alpha k}+i\sum_\nu \frac{\partial B_{\alpha\nu}^{(1)} }{\partial z_0}\phi_{\nu k} -\sum_\nu (\tb_{\alpha}-\tb_{\nu})B_{\alpha \nu}^{(1)}\phi_{\nu k}=-
  \frac{\partial A_\alpha}{\partial y_1}\frac{\partial \phi_{\alpha k}}{\partial y_0},
  \\
  \label{mu2order_12}
  i\frac{\partial A_{\beta}}{\partial z_1}\phi_{\beta k}+i\sum_\nu \frac{\partial B_{\beta\nu}^{(1)} }{\partial z_0}\phi_{\nu k} -\sum_\nu (\tb_{\beta}-\tb_{\nu})B_{\beta \nu}^{(1)}\phi_{\nu k}=-\frac{\partial A_\beta}{\partial x_1}\frac{\partial \phi_{\beta k}}{\partial y_0}.
\end{eqnarray}
 \end{subequations}
 \end{widetext}
 
 Now we apply $(\phi_{\alpha k},\cdot)$ to (\ref{mu2order_11}) to obtain
\begin{eqnarray}
\label{aux_A}
 i\frac{\partial A_{\alpha}}{\partial z_1}+i\left(\psi_{\alpha k},\frac{1}{i}\frac{\partial}{\partial y_0}\psi_{\alpha k}\right)\frac{\partial A_{\alpha}}{\partial y_1}+\frac{\partial B_{\alpha\alpha}^{(1)} }{\partial z_0}=0,
\end{eqnarray}
where we have used  
\begin{eqnarray}
\left(\phi_{\nu k},\frac{1}{i}\frac{\partial}{\partial y_0}\phi_{\alpha k}\right)\equiv \left(\psi_{\nu k},\frac{1}{i}\frac{\partial}{\partial y_0}\psi_{\alpha k}\right).
\end{eqnarray}
Taking into account that all terms in (\ref{aux_A}) are $T-$periodic with respect to $z_0$,  we average (\ref{aux_A}) over the period $T$ to obtain
\begin{eqnarray}
\label{Afirst1}
\frac{\partial A_{\alpha}}{\partial z_1}+v_{\alpha}\frac{\partial A_{\alpha}}{\partial y_1}=0,
\end{eqnarray}
where the group velocity $v_{\alpha}$ is given by (\ref{omega1}). Analogously from (\ref{mu2order_12}) we obtain  
\begin{eqnarray}
\label{Afirst2}
\frac{\partial A_{\beta}}{\partial z_1}+v_{\beta}\frac{\partial A_{\beta}}{\partial y_1}=0.
\end{eqnarray}

An important property, is used in the main text, is the equality of the group velocities of the chosen modes, i.e., 
\begin{align}
\label{group-velocity}
    v_\alpha=v_\beta=v.
\end{align}
This means that both $A_\alpha$ and $A_\beta$ depend on the ``fast variables'' $z_1$ and $y_1$ only through the combination $Y=y_1-v z_1$: 
\begin{eqnarray}
\label{AX}
A_{\alpha,\beta}\equiv A_{\alpha,\beta}(Y; z_2,x_2).
\end{eqnarray}
We also obtain  
\begin{eqnarray}
\label{Ax1}
B_{\alpha\alpha}^{(1)} 
=-i\frac{\partial A_\alpha}{\partial y_1}c_{\alpha\alpha}^{(1)}\;, \quad B_{\beta\beta}^{(1)} 
=-i\frac{\partial A_\beta}{\partial y_1}c_{\beta\beta}^{(1)}\,,
\end{eqnarray}
where $c_{\nu\nu}^{(1)}$ is defined in  (\ref{b1}). Importantly, at this stage we assume that the modes are non-resonant, i.e., no zero denominators appear in (\ref{b1}).

For $\nu\neq \alpha$ we apply $(\phi_{\nu k},\cdot)$ to (\ref{mu2order_11}) and obtain [cf. (\ref{ode_b1})]
\begin{eqnarray}
 i\frac{\partial B_{\alpha\nu}^{(1)} }{\partial z_0}-(\tb_{\alpha}-\tb_{\nu})B_{\alpha\nu}^{(1)}=-i\frac{\partial A}{\partial y_1}h_{\alpha\nu}^*(z_0).
\end{eqnarray}
Its $T-$ periodic solution is found as  
\begin{eqnarray}
\label{B1_fin}
B_{\alpha\nu}^{(1)}(z_0)= -  i\frac{\partial A_\alpha}{\partial y_1}c_{\alpha \nu}^{(1)}\,,
\end{eqnarray}
where $c_{\alpha \nu}^{(1)}$ is defined in (\ref{b1_fin}). Similarly, for the $\beta-$component we obtain
\begin{eqnarray}
\label{B1_fin_b}
B_{\beta\nu}^{(1)}(z_0)= -  i\frac{\partial A_\beta}{\partial y_1}c_{\beta \nu}^{(1)}\,.
\end{eqnarray}

 \subsubsection{Order $\mu^3$}
 
   Turning to the equations of the $\mu^3$ order we write them already separated for the terms $\propto e^{i\tb_{\alpha} z}$ and $\propto e^{i\tb_{\beta} z}$, where all entries $\propto e^{\pm 3i\tb_{\alpha} z}$ and $\propto e^{\pm 3i\tb_{\beta} z}$ are dropped:
    \begin{widetext}
 \begin{align}
 \label{mu3order-1}
& i\frac{\partial A_{\alpha}}{\partial z_2}\phi_{\alpha k}+i\sum_\nu \frac{\partial B_{\alpha\nu}^{(2)} }{\partial z_0}\phi_{\nu k}+i\sum_\nu \frac{\partial B_{\alpha\nu}^{(1)} }{\partial z_1}\phi_{\nu k} +i\sum_\nu B_{\alpha\nu}^{(2)} \frac{\partial \phi_{\nu k}}{\partial z_0} -\tb_{\alpha}\sum_\nu B_{\alpha\nu}^{(2)} \phi_{\nu k}
 \nonumber \\
& =\tH_0\sum_\nu B_{\alpha \nu}^{(2)} \phi_{\nu k}+\tH_1 \sum_\nu B_{\alpha \nu}^{(1)} \phi_{\nu k}
  +H_2A_\alpha \phi_{\alpha k}-|A_\alpha|^2A_\alpha|\phi_{\alpha k}|^2\phi_{\alpha k}-2|A_\beta|^2A_\alpha|\phi_{\beta k}|^2\phi_{\alpha k}\,,
  \\
   \label{mu3order-2}
  & i\frac{\partial A_{\beta}}{\partial z_2}\phi_{\alpha k}+i\sum_\nu \frac{\partial B_{\beta\nu}^{(2)} }{\partial z_0}\phi_{\nu k}+i\sum_\nu \frac{\partial B_{\beta\nu}^{(1)} }{\partial z_1}\phi_{\nu k} +i\sum_\nu B_{\beta\nu}^{(2)} \frac{\partial \phi_{\nu k}}{\partial z_0} -\tb_{\beta}\sum_\nu B_{\beta\nu}^{(2)} \phi_{\nu k}
 \nonumber \\
& =\tH_0\sum_\nu B_{\beta\nu}^{(2)} \phi_{\nu k}+\tH^{(1)} \sum_\nu B_{\alpha \nu}^{(1)} \phi_{\nu k}
  +\tH^{(2)}A_\beta \phi_{\beta k}-|A_\beta|^2A_\beta|\phi_{\beta k}|^2\phi_{\beta k}-2|A_\alpha|^2A_\beta|\phi_{\alpha k}|^2\phi_{\beta k}\,.
 \end{align}
 Projecting  (\ref{mu3order-1}) and  (\ref{mu3order-2}) to $\phi_{\alpha k}$ and $\phi_{\beta k}$ respectively we obtain
\begin{align}
\label{mu3order-11}
i\frac{\partial A_{\alpha}}{\partial z_2}+i\frac{\partial B_{\alpha\alpha}^{(2)} }{\partial z_0}+i\frac{\partial B_{\alpha\alpha}^{(1)} }{\partial z_1}= -\sum_\nu \frac{\partial B_{\alpha\nu}^{(1)}}{\partial y_1}  \left(\phi_{\alpha k},\frac{\partial \phi_{\alpha k} }{\partial y_0}\right)-\frac{1}{2}\frac{\partial^2 A_\alpha}{\partial y_1^2}- \frac{\partial A_\alpha}{\partial y_2}\left(\phi_{\alpha k},\frac{\partial \phi_{\alpha k} }{\partial y_0}\right)
\nonumber \\-|A_\alpha|^2A_\alpha(\phi_{\alpha k},|\phi_{\alpha k}|^2\phi_{\alpha k})-2|A_\beta|^2A_\alpha(\phi_{\alpha k},|\phi_{\beta k}|^2\phi_{\alpha k}),
\\
\label{mu3order-12}
i\frac{\partial A_{\beta}}{\partial z_2}+i\frac{\partial B_{\beta\beta}^{(2)} }{\partial z_0}+i\frac{\partial B_{\beta\beta}^{(1)} }{\partial z_1}= -\sum_\nu \frac{\partial B_{\beta\nu}^{(1)}}{\partial y_1}  \left(\phi_{\beta k},\frac{\partial \phi_{\beta k} }{\partial y_0}\right)-\frac{1}{2}\frac{\partial^2 A_\beta}{\partial y_1^2}- \frac{\partial A_\beta}{\partial y_2}\left(\phi_{\beta k},\frac{\partial \phi_{\beta k} }{\partial y_0}\right)
\nonumber \\-|A_\beta|^2A_\beta(\phi_{\beta k},|\phi_{\beta k}|^2\phi_{\beta k})-2|A_\alpha|^2A_\beta(\phi_{\beta k},|\phi_{\alpha k}|^2\phi_{\beta k}).
\end{align}
It follows from (\ref{B1_fin}), (\ref{Afirst1}), (\ref{Afirst2}) and (\ref{group-velocity}) that 
\begin{eqnarray}
i \frac{\partial B_{\alpha\nu}^{(1)} }{\partial z_1} =v \frac{\partial^2 A_\alpha}{\partial y_1^2}c_{\alpha\nu}^{(1)}, \qquad i \frac{\partial B_{\beta\nu}^{(1)} }{\partial z_1} =v \frac{\partial^2 A_\beta}{\partial y_1^2}c_{\beta\nu}^{(1)}\,.
\end{eqnarray}
Thus (\ref{mu2order_11}) and (\ref{mu2order_12}) are rewritten as
\begin{align} 
\label{eq:aux1}
i\frac{\partial A_{\alpha}}{\partial z_2}+iv\frac{\partial A_{\alpha}}{\partial y_2}+\frac{1}{2}\frac{\partial^2 A_\alpha}{\partial y_1^2}+i\sum_{\nu\neq \alpha} h_{\alpha\nu}\frac{\partial B_{\alpha\nu}^{(1)} }{\partial y_1}+|A_\alpha|^2A_\alpha(\phi_{\alpha k},|\phi_{\alpha k}|^2\phi_{\alpha k})   +2|A_\beta|^2A_\alpha(\phi_{\alpha k},|\phi_{\beta k}|^2\phi_{\alpha k})    
=  -i\frac{\partial B_{\alpha\alpha}^{(2)}}{\partial z_0},
\\
\label{eq:aux2}
i\frac{\partial A_{\beta}}{\partial z_2}+iv\frac{\partial A_{\beta}}{\partial y_2}+\frac{1}{2}\frac{\partial^2 A_\beta}{\partial y_1^2}+i\sum_{\nu\neq \beta} h_{\beta\nu}\frac{\partial B_{\beta\nu}^{(1)} }{\partial y_1}+|A_\beta|^2A_\beta(\phi_{\beta k},|\phi_{\beta k}|^2\phi_{\beta k})   +2|A_\beta|^2A_\alpha(\phi_{\beta k},|\phi_{\alpha k}|^2\phi_{\beta k})    
=  -i\frac{\partial B_{\beta\beta}^{(2)}}{\partial z_0}.
\end{align}
\end{widetext}

Since all terms in this equations are either $z_0-$independent or $T-$periodic 
we  average over the period. The last term in (\ref{eq:aux1}) and  (\ref{eq:aux2}) vanish because of the periodicity, while  using (\ref{B1_fin}) and (\ref{omega2}) we obtain  
\begin{align}
&\frac{1}{2}\frac{\partial^2 A_\alpha}{\partial y_1^2}+i\left\langle \sum_{\nu\neq \alpha} h_{\alpha\nu}\frac{\partial B_{\alpha \nu}^{(1)} }{\partial y_1} \right\rangle_T
\nonumber \\
&=\frac{1}{2}\frac{\partial^2 A_\alpha}{\partial y_1^2}\left(1+2\sum_{\nu\neq \alpha}\left\langle h_{\alpha\nu}c_{_\alpha\nu}^{(1)}  \right\rangle_T\right)
=-\frac{\tb_{\alpha}^{\prime\prime}}{2}\frac{\partial^2 A_\alpha}{\partial y_1^2}.
\end{align}
Similar relation holds for $\alpha$ replaced by $\beta$.

Looking for the envelopes  $A_{\alpha}$ and $A_{\beta}$ which are $y_2-$independent, we obtain two nonlinearly coupled NLS equations. Setting the formal small parameter $\mu$ to be one, i.e., returning to non-scaled physical variables we obtain 
\begin{subequations}
\begin{align}
\label{coupled1}
i\frac{\partial A_{\alpha}}{\partial z}-\frac{\tb_{\alpha}^{\prime\prime}}{2}
\frac{\partial^2A_\alpha}{\partial Y^2}+\chi_{\alpha} |A_\alpha|^2A_\alpha+2\chi_{\rm x} |A_\beta|^2A_\alpha=0,
\\
\label{coupled2}
i\frac{\partial A_{\beta}}{\partial z}-\frac{\tb_{\beta}^{\prime\prime}}{2}\frac{\partial^2A_\beta}{\partial Y^2}+\chi_{\beta} |A_\beta|^2A_\beta+2\chi_{\rm x} |A_\alpha|^2A_\beta=0,
\end{align}
\end{subequations}
where the nonlinearity coefficients are given by
\begin{align}
\chi_\alpha&=\langle (\psi_{\alpha k},|\psi_{\alpha k}|^2\psi_{\alpha k})\rangle_T\,,
\\
\chi_\beta&=\langle (\psi_{\beta k},|\psi_{\beta k}|^2\psi_{\beta k})\rangle_T\,,
\\
\chi_{\rm x}&=\langle (\psi_{\alpha k},|\psi_{\beta k}|^2\psi_{\alpha k})\rangle_T=\langle (\psi_{\beta k},|\psi_{\alpha k}|^2\psi_{\beta k})\rangle_T\,.
\end{align}
These are equations~(2) from the main text, where $\tb_\nu^{\prime\prime}$ is replaced by $b_\nu^{\prime\prime}$ referring to the linear spectrum.

 \section{Linear stability analysis of system~(2)}

To perform linear stability analysis in the frames of the envelope equation~(2) from the main text we substitute into it the perturbed solution $A_\nu=(w_\nu+\mu_\nu e^{\delta z}+\eta_\nu^{*} e^{\delta^* z})e^{i b_\nu^\textrm{nl}z}$, where $\mu_\nu, \eta_\nu$ are small perturbations, $\delta$ is the complex perturbation growth rate, and linearize resulting system. This yields linear eigenvalue problem:

\begin{subequations}
\begin{widetext}
\begin{align}
i\delta \mu_{\alpha,\beta}=+\frac{b_{\alpha,\beta}^{\prime\prime}}{2}\frac{\partial^2 \mu_{\alpha,\beta}}{\partial Y^2}-\chi_{\alpha,\beta} (2\mu_{\alpha,\beta}+\eta_{\alpha,\beta})w_{\alpha,\beta}^2-2\chi_\textrm{x}[w_{\beta,\alpha}^2 \mu_{\alpha,\beta}+w_{\alpha}w_{\beta}(\mu_{\beta,\alpha}+\eta_{\beta,\alpha})]+b_{\alpha,\beta}^\textrm{nl}\mu_{\alpha,\beta}\,,
\\
i\delta \eta_{\alpha,\beta}=-\frac{b_{\alpha,\beta}^{\prime\prime}}{2}\frac{\partial^2 \eta_{\alpha,\beta}}{\partial Y^2}+\chi_{\alpha,\beta} (2\eta_{\alpha,\beta}+\mu_{\alpha,\beta})w_{\alpha,\beta}^2+2\chi_\textrm{x}[w_{\beta,\alpha}^2 \eta_{\alpha,\beta}+w_{\alpha}w_{\beta}(\eta_{\beta,\alpha}+\mu_{\beta,\alpha})]-b_{\alpha,\beta}^\textrm{nl}\eta_{\alpha,\beta}\,.
\end{align}
\end{widetext}
\end{subequations}

Stationary envelopes $w_{\alpha,\beta}$ that enter linear eigenvalue problem [Eqs.~(73a) and~(73b)] can be found using Newton method from the following system of nonlinearly coupled ordinary differential equations [also obtained from Eq.~(2) of the main text]:

\begin{subequations}
\begin{align}
\label{coupled3}
b_\alpha^\textrm{nl}w_{\alpha}= -\frac{b_{\alpha}^{\prime\prime}}{2}
\frac{\partial^2w_\alpha}{\partial Y^2}+\chi_{\alpha} w_\alpha^3+2\chi_{\rm x} w_\beta^2w_\alpha,
\\
\label{coupled4}
b_\beta^\textrm{nl}w_{\beta}=-\frac{b_{\beta}^{\prime\prime}}{2}\frac{\partial^2w_\beta}{\partial Y^2}+\chi_{\beta} w_\beta^3+2\chi_{\rm x} w_\alpha^2w_\beta.
\end{align}
\end{subequations}

Linear eigenvalue problem [Eqs. (73a) and (73b)] was solved using standard eigenvalue solver to obtain the dependence of perturbation growth rate $\delta_\textrm{re}=\textrm{Re}(\delta)$ on nonlinear propagation constant shifts $b_\alpha^\textrm{nl}$ and $b_\beta ^\textrm{nl}$ of two soliton components. We identified the most unstable perturbation mode with largest growth rate and plotted it as a function of $b_\alpha^\textrm{nl}$ for several $b_\beta ^\textrm{nl}$ values in Fig. 3(d) from the main text. One can see, that for nonlinear phase shifts used in the paper (they should be sufficiently small to ensure that the envelope covers many $y$-periods of the array) growth rate for the most unstable perturbation eigenmode is typically very low. Thus, for soliton shown in Fig. 3(b) one has $\delta_\textrm{re}=0.00044$. The characteristic propagation distance at which such instability may develop can be estimated as $1/\delta_\textrm{re}$ and for all cases considered it exceeds helix period $T$ at least by two orders of magnitude that implies metastability (very long-living character) of the obtained dipole solitons. Due to their metastability, one can observe long-range propagation of dipole solitons along the edge of the insulator without breakup into sets of fundamental solitons. Notice that perturbation growth rate vanishes close to the left border of the existence domain of vector solitons, when $b_\alpha^\textrm{nl} \to b_\alpha^\textrm{low}$ , but so does also the dipole component of soliton. Thus, in the paper the optimal situation was chosen, when this component is still considerable in comparison with other bell-shaped component, and at the same time, growth rate $\delta_\textrm{re}$   remains very small. The analysis described above guarantees metastability of the one-dimensional envelopes of vector topological edge solitons.

\section{Evolution of peak amplitudes in linear and nonlinear regimes}

The fact that dipole topological solitons are indeed the objects, sustained by the nonlinearity of the material, becomes especially obvious from comparison of evolution of peak amplitudes of the wavepackets in nonlinear and linear cases. Such a comparison is presented in Fig.~\ref{Fig1} below that shows the dependence of the maximal amplitude of the wavepacket propagated in scalar model~(1) from the main text [Fig.~\ref{Fig1}(a), $a=\textrm{max}|\psi|$] and wavepacket, whose components evolve in accordance with vector model~(3) [Fig.~\ref{Fig1}(b), $a_{\alpha,\beta}=\textrm{max}|\psi_{\alpha,\beta}|$]. The amplitude in nonlinear medium is shown with black (in scalar case) or black and red (in vector case) lines. Note that fast oscillations with period $T$ clearly visible in the plots reflect the underlying Floquet nature of the obtained solitons. While in nonlinear case peak amplitude does not decrease notably over considerable distance shown, in linear versions of the above mentioned models, where nonlinearity was deliberately switched off (see green lines), peak amplitude rapidly drops down reflecting strong diffraction broadening observed in Figs.~\ref{Fig4}(d) and~\ref{Fig5}(d) from the main text.

\begin{figure}[t]
\centering
\includegraphics[width=\linewidth]{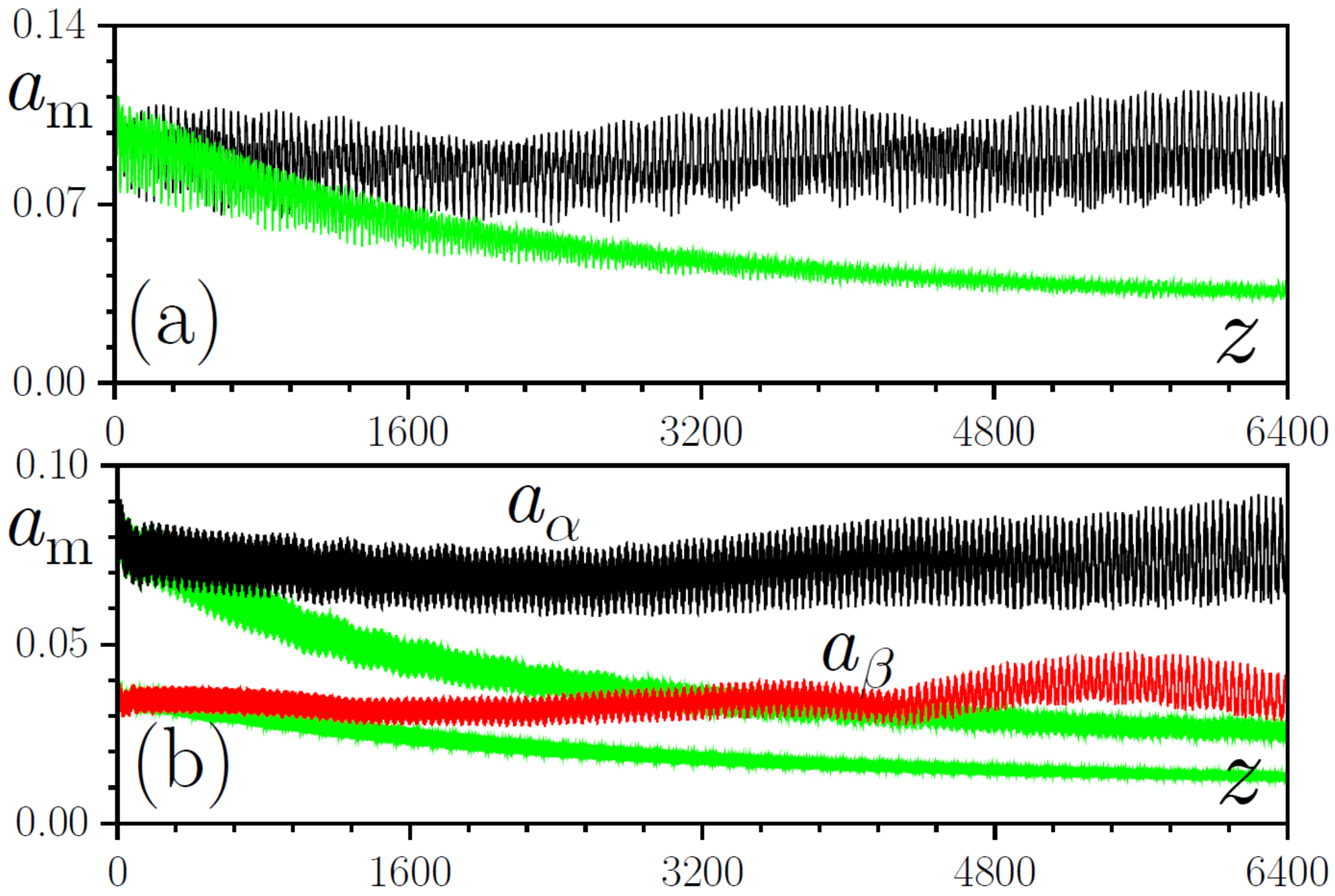}
\caption{Peak amplitude of the dipole soliton versus distance in (a) scalar and (b) vector models corresponding to the dynamics shown in Figs.~\ref{Fig4} and~\ref{Fig5} from the main text. Black and red curves - peak amplitude in nonlinear regime, green curves – peak amplitude in linear regime.}
\label{figure1}
\end{figure}

\end{document}